\documentstyle[12pt]{article}

\begin{document}
\begin{titlepage}
\hskip 12cm \vbox{\hbox{BUDKERINP/98-54}\hbox{UNICAL-TP 98/3}
\hbox{July 1998}}
\vskip 0.3cm
\centerline{\bf THE GENERALIZED NON-FORWARD BFKL EQUATION}
\centerline{\bf AND THE ``BOOTSTRAP'' CONDITION} 
\centerline{\bf FOR THE GLUON REGGEIZATION IN THE NLLA$^{~\ast}$}
\vskip 1.0cm
\centerline{  V.S. Fadin$^{a~\dagger}$, R. Fiore$^{b~\ddagger}$}
\centerline{\sl $^{a}$ Budker Institute for Nuclear Physics,}
\centerline{\sl Novosibirsk State University, 630090
Novosibirsk, Russia}
\vskip 0,5cm
\centerline{\sl $^{b}$ Dipartimento di Fisica, Universit\`a della
Calabria,}
\centerline{\sl Istituto Nazionale di Fisica Nucleare, Gruppo
collegato di Cosenza,}
\centerline{\sl Arcavacata di Rende, I-87036 Cosenza, Italy}
\vskip 1cm
\begin{abstract}
The generalization of the BFKL equation for the case of non-forward 
scattering is considered. The kernel of the generalized equation in the 
next-to-leading approximation is expressed in terms of the gluon Regge 
trajectory and the effective vertices for particle production in Reggeon 
collisions. The ``bootstrap'' equations for the gluon Reggeization are 
presented.
\end{abstract}
\vskip .5cm
\hrule
\vskip.3cm
\noindent

\noindent
$^{\ast}${\it Work supported in part by the Ministero italiano
dell'Universit\`a e della Ricerca Scientifica e Tecnologica, in part
by INTAS and in part by the Russian Fund of Basic Researches.}
\vfill
$ \begin{array}{ll}
^{\dagger}\mbox{{\it email address:}} &
 \mbox{FADIN~@INP.NSK.SU}\\
\end{array}
$

$ \begin{array}{ll}
^{\ddagger}\mbox{{\it email address:}} &
  \mbox{FIORE~@FIS.UNICAL.IT}
\end{array}
$
\vfill
\vskip .1cm
\vfill
\end{titlepage}
\eject
\textheight 210mm \topmargin 2mm \baselineskip=24pt

\vskip 0.5cm

\section{Introduction}

One of the remarkable properties of QCD\ is the Reggeization of elementary
particles. Contrary to QED, where the electron does Reggeize in perturbation
theory \cite{GGLMZ}, but the photon remains elementary \cite{Man}, in QCD
the gluon [3,4] does Reggeize as well as the quark \cite{FaSh}. The gluon
Reggeization plays the key role in the derivation of the BFKL equation \cite
{BFKL} for the cross sections at high C.M.S. energy $\sqrt{s}$ in
perturbative QCD. This equation is very important for the theory of high
energy processes. It is used \cite{NZZ} together with the DGLAP equation 
\cite{DGLAP} for the description of deep inelastic scattering 
processes at a small value of the Bjorken variable $x$. The equation was
derived \cite{BFKL} in the leading logarithm approximation (LLA) more than
twenty years ago, and recently the calculation of radiative corrections
[9-15] to the kernel of the equation was completed and the equation in the
next-to-leading logarithmic approximation (NLLA) was obtained 
\cite{FaLi98}.

The famous BFKL equation is a particular case of the equation for the 
$t$-channel partial waves of the elastic amplitudes \cite{BFKL} for the 
forward 
 scattering, 
i.e. $t=0$ and vacuum quantum numbers in the $t$-channel. Evidently, it is 
very important to obtain in the NLLA not only the equation for this particular
case, but  the equation for the non-forward scattering as well. Besides the 
fact that the last equation is much more general, it permits to check in 
the NLLA the gluon Reggeization, the base of the whole program of the 
calculation of the radiative corrections formulated in Ref.~\cite{FaLi89} 
and fulfilled in Refs.~[9-15]. Remind that in the LLA the Reggeization was 
noticed in the first several orders of the perturbation theory. After that, 
assuming that it is correct in all orders, the equation for the $t$-channel 
partial waves of the elastic scattering amplitudes  \cite{BFKL} was derived.
It is clear that, for the gluon quantum numbers in the $t$-channel, 
the solution of this equation must reproduce the gluon Reggeization, 
as it was explicitly demonstrated in 
Ref.~\cite{BFKL}. This ``bootstrap'' supports the idea of the
Reggeization in such a strong way that practically no doubts remains that
it is correct. Nevertheless, strictly speaking, the ``bootstrap'' cannot be
considered as a rigorous proof. Therefore, such
a proof was specially constructed in Ref.~\cite{BaLiFa}. In the NLLA till 
now we have only the simple check of the Reggeization in
the first three orders of perturbation theory in $\alpha _{S}$ \cite{FFKQ}.

In this paper we present the representation for the scattering amplitudes 
in QCD at high energy $\sqrt{s}$ and fixed momentum transfer $\sqrt{-t}$ in 
the NLLA in terms of the impact factors of the scattered particles and the 
Green function for the Reggeized gluon scattering. The representation is 
obtained on the base of the gluon Reggeization. The impact factors and the 
kernel of the equation for the Green function 
are expressed in terms of the gluon Regge trajectory  and the effective
vertices for Reggeon-Reggeon and Reggeon-particle interaction. The 
requirement of the selfconsistency  leads to the ``bootstrap'' equations 
for the gluon Reggeization.

In the next Section we discuss the meaning of the gluon Reggeization. In
Section 3 we show the representation of the scattering amplitudes in
terms of the impact factors and the Green function. In Section 4 the
"bootstrap" equations are derived. The summary is given in Section 5.

\bigskip

\section{The gluon Reggeization in QCD}

The notion ``Reggeization'' of elementary particles in perturbation theory
is usually related to the absence of non analytic terms in the complex
angular momentum plane [1-3]. We use this notion in a much stronger sense.
Talking about the gluon Reggeization in QCD we mean not only the existence
of the Reggeon with gluon quantum numbers, negative signature and trajectory 
\begin{equation}
j\left( t\right) =1+\omega(t)  
\label{z1}
\end{equation}
passing through $1$ at $t=0$. We mean also that in each order of
perturbation theory this Reggeon gives the leading contribution to the 
amplitudes of the processes at large relative energies of the participating
particles and fixed (i.e. not increasing with $s$) momentum transfers.

Let us explain this in more details. Consider the elastic scattering process 
$A+B$ $\rightarrow A^{\prime }+B^{\prime }$ at large $s$ and fixed $t$:
\begin{equation}
s=(p_{A}+p_{B})^{2}\rightarrow \infty \ ,\mbox{ \ \ \ \ \ \ \ }t=q^{2}
\mbox{\ \ fixed , \ \ \ \ \ \ }q=p_{A}-p_{B}~.  
\label{z2}
\end{equation}
For the sake of brevity, the term ``gluon Reggeization'' used by us means 
that the elastic scattering amplitude with the gluon quantum numbers in the 
$t$-channel has the Regge form
\begin{equation}
({\cal A}_{8})_{AB}^{A^{\prime }B^{\prime }}=\Gamma _{A^{\prime }A}^{c}\left[
\left( \frac{-s}{-t}\right) ^{j(t)}-\left( \frac{+s}{-t}\right) ^{j(t)}%
\right] \Gamma _{B^{\prime }B\mbox{ \ \ }}^{c} .  
\label{z3}
\end{equation}
Here $c$ is a color index and $\Gamma _{P^{\prime }P}^{c}$ are the
particle-particle-Reggeon (PPR) vertices which do not depend on $s$. Notice
that the form (\ref{z3}) represents correctly the analytical structure of
the scattering amplitude, which is quite simple in the elastic case. 
In the derivation of the BFKL equation it is assumed that this form 
is valid in the NLLA as well as in the LLA.

Together with the form (\ref{z3}) of the elastic amplitude the derivation of
the BFKL equation in the LLA and NLLA is based on the Reggeized form of
production amplitudes in the multi-Regge kinematics (MRK). For the
production of $n$ particles with momenta $k_{i}$, $i=1\div n$, in the
process $A+B\rightarrow \tilde{A}+\tilde{B}+n$ this kinematics implies
that the invariant masses of any pair of produced particles are large and
all the transferred momenta are fixed (not increasing with $s$). More
definitely, let us put $p_{\tilde{A}}\equiv k_{0}$, $p_{\tilde{B}}\equiv k_{%
n+1}$ and introduce the Sudakov decomposition
\begin{equation}
k_{i}=\beta _{i}p_{1}+\alpha _{i}p_{2}+k_{i\perp} \ ,\mbox{ \ \ \ \ \ \ \ \ }
s\alpha _{i}\beta _{i}=k_{i}^{2}-k_{i\perp }^{2}=
k_{i}^{2}+\vec{k}_{i}^{~2}~,  
\label{z4}
\end{equation}
where $p_{1,2}$ are the light cone momenta such that
\begin{equation}
p_{A}=p_{1}+\frac{m_{A}^{2}}{s}p_{2} \ , \mbox{ \ \ \ \ \ \ }p_{B}=p_{2}+%
\frac{m_{B}^{2}}{s}p_{1} \ , \mbox{ \ \ \ \ \ \ }2p_{1}p_{2}=s  
\label{z5}
\end{equation}
(we admit all particles to have non zero masses, reserving the possibility
to consider each of them as a compound state or as a group of particles) and
the vector sign is used for the transverse components. Then in the MRK we
have
\begin{displaymath}
\frac{\vec{p}_{\tilde{A}}^{~2}+m_{\tilde{A}}^{2}}{s} \approx \alpha _{0}
\ll \alpha _{1}\dots \ll \alpha _{n}\ll \alpha _{n+1}\approx 1 \ ,  
\end{displaymath}
\begin{equation}
\frac{\vec{p}_{\tilde{B}}^{~2}+m_{\tilde{B}}^{2}}{s} \approx \beta _{n+1}
\ll \beta _{n}\dots \ll \beta _{1}\ll \beta _{0}\approx 1 \ .
\label{z6}
\end{equation}
Due to Eqs.~(\ref{z4})-(\ref{z6}) the squared invariant masses
\begin{equation}
s_{i}=(k_{i-1}+k_{i})^{2}\approx s\beta _{i-1}\alpha _{i}=
\frac{\beta _{i-1}}{\beta _{i}}(k_{i}^{2}+\vec{k}_{i}^{2})
\label{z7}
\end{equation}
are large compared with the squared transverse momenta of the produced 
particles, which have the order of the squared momentum transfers:
\begin{equation}
s_{i}\gg \vec{k}_{i}^{2}\sim \mid t_{i}\mid =\mid q_{i}^{2}\mid \mbox{\ ,}
\label{z8}
\end{equation}
where
\begin{displaymath}
q_{i}=p_{A}-\sum_{j=0}^{i-1}k_{j} =
-\left(p_{B}-\sum_{j=i}^{n+1}k_{j}\right)
\approx \beta _{i}p_{1}-\alpha _{i-1}p_{2}-
\sum\limits_{j=0}^{i-1}k_{j\perp }\mbox{\ ,}
\end{displaymath}
\begin{equation}
t_{i}=q_{i}^{2}\approx q_{i\perp }^{2}=-\vec{q}_{i}^{~2}\mbox{\ ,}
\label{z9}
\end{equation}
and the product of $s_{i}$ is proportional to $s$:
\begin{equation}
\prod\limits_{i=1}^{n+1}s_{i}=s\prod\limits_{i=1}^{n}(k_{i}^{2}+
\vec{k}_{i}^{2})\mbox{\ .}
\label{z10}
\end{equation} 
It is necessary to remind that, contrary to the elastic amplitude, the 
production amplitudes have a complicated analytical structure (see, for
instance, Refs.~[10,18]). Fortunately, only the real parts of these amplitudes
are used in the derivation of the BFKL equation in the NLLA as well as in
the LLA. The term ``gluon Reggeization'' used by us means that the real parts 
of the production amplitudes in the MRK have a simple factorized form and can 
be presented as 
\begin{displaymath}
{\cal A}_{AB}^{\tilde{A}\tilde{B}+n} = 2s\Gamma _{\tilde{A}A}^{c_{1}}
\left[\prod_{i=1}^{n}\frac{1}{t_{i}}
\gamma_{c_{i}c_{i+1}}^{P_{i}}(q_{i},q_{i+1})\left( \frac{s_{i}}
{\sqrt{\vec{k}_{i-1}^{2}\vec{k}_{i}^{2}}}\right)^{\omega (t_{i})}\right]  
\end{displaymath}
\begin{equation}
\times \frac{1}{t_{n+1}}\left( \frac{s_{n+1}}{\sqrt{\vec{k}_{n}^{2}
\vec{k}_{n+1}^{2}}}\right) ^{\omega (t_{n+1})}
\Gamma _{\tilde{B}B\mbox{ \ \ }}^{c_{n+1}}\mbox{\ .}
\label{z11}
\end{equation}
Here $\gamma _{c_{i}c_{i+1}}^{P_{i}}(q_{i},q_{i+1})$ are the 
so-called Reggeon-Reggeon-particle (RRP) vertices,
i.e. the effective vertices for the production of the particles $P_i$ with
momenta $k_{i}$=$q_{i}-q_{i+1}$ in the collision of the
Reggeons with momenta $q_{i}$ and $q_{i+1}$ and colour indices
$c_{i}$ and $c_{i+1}$. Pay attention that we have taken
definite energy scales in the Regge factors in Eq.~(\ref{z11}) as well as 
in
Eq.~(\ref{z3}). In principle, we could take an arbitrary scale $s_{R}$; in
this case the PPR and RRP vertices would become dependent on $s_{R}$. Of
course, physical results do not depend on the scale.

In the LLA only one particle can be produced in the RRP vertex, and since
our Reggeons are Reggeized gluons, this particle can be only a gluon. The
situation is quite different in the NLLA. In this case we have to consider
the so-called quasi-multi-Regge kinematics (QMRK) \cite{FaLi89}, where any
(but only one) pair\ of the produced particles has a fixed (not increasing
with $s$) invariant mass. We can treat this kinematics using the effective
vertices $\gamma _{c_{i}c_{i+1}}^{G_{1}G_{2}}(q_{i},q_{%
i+1})$ [9,13] and $\gamma _{c_{i}c_{i+1}}^{Q\bar{Q}}(q_{%
i},q_{i+1})$ [13,15] for the production of two gluons and a
quark-antiquark pair respectively in Reggeon-Reggeon  collisions, as 
well as the effective vertices $\Gamma_{P^{\ast }P}^{c}$ for the production 
of the ``excited'' state (containing
an extra particle) in the fragmentation region of the particle $P$ in the
process of scattering of this particle off the Reggeon. Introducing these
vertices the production amplitudes of $n+1$ particles in the QMRK are 
given
by Eq.~(\ref{z11}) with one of the vertices $\gamma _{c_{i}c_{i%
+1}}^{P_{i}}$ or $\Gamma _{\tilde{P}P}^{c}$ substituted with the
vertices $\gamma _{c_{i}c_{i+1}}^{P_{1}P_{2}}$ or $\Gamma
_{P^{\ast }P}^{c}$ respectively.

Since in the limit of large invariant masses of all pairs of the final
state particles the QMRK amplitudes must turn into the MRK ones, in this
limit the effective vertices $\gamma _{c_{i}c_{i+1}}^{G_{1}G_{2}}$ 
satisfy the factorization properties
\begin{equation}
\gamma _{c_{i}c_{i+1}}^{G_{1}G_{2}}(q_{i},q_{i%
+1})=\gamma _{c_{i}c}^{G_{1}}\left( q_{i},q_{i%
}-l_{1}\right) \frac{1}{\left( q_{i}-l_{1}\right) ^{2}}\gamma _{cc_{%
i+1}}^{G_{2}}\left( q_{i}-l_{1},q_{i+1}\right)
\label{z12}
\end{equation}
at $\left( p_{B}l_{2}\right) \ll \left( p_{B}l_{1}\right) $, $\left(
p_{A}l_{1}\right) \ll \left( p_{A}l_{2}\right)$, where $l_{1,2}$ are the
momenta of the produced gluons. The vertices $\Gamma _{P^{\ast }P}^{c}$ in 
the case in which the ``excited'' state $P^{\ast }$ contains the gluon, i.e.
$P^{\ast }=G\tilde{P}$, have the property
\begin{equation}
\Gamma _{G\tilde{P}P}^{c}\left( q\right) \simeq
\Gamma _{\tilde{P}P}^{\tilde{c}}%
\frac{1}{\left( q+l\right) ^{2}}\gamma _{\tilde{c}c}^{G}\left( q+l,q\right)
\label{z13}
\end{equation}
at $\left( p_{P}l\right) \gg \left( p_{P}p_{\tilde{P}}\right) $, $l$ being
the gluon momentum and $q=p_{P}-p_{\tilde{P}}$.

The BFKL equation is straightforwardly obtained \cite{BFKL} if the 
amplitudes (\ref{z3}) and (\ref{z11}) are used in the unitarity relation for 
the $s$-channel imaginary part of the elastic scattering amplitude. 
Remind that the representations (\ref{z3}) and (\ref{z11}) for the 
amplitudes with the gluon quantum numbers in the $t_{i}$-channels were 
rigorously proved \cite{BaLiFa} in the LLA.

\bigskip

\section{The generalized BFKL equation in the NLLA}

Decomposing the elastic scattering amplitudes ${\cal A}_{AB}^{A^{^{\prime
}}B^{^{\prime }}}$in the parts with definite irreducible representation 
${\cal R}$ of the colour group in the $t$-channel:
\begin{equation}
{\cal A}_{AB}^{A^{\prime }B^{\prime }} =\sum\limits_{{\cal R}}
\left({\cal A}_{{\cal R}}\right)_{AB}^{A^{\prime }B^{\prime }}\mbox{\ ,}
\label{z14}
\end{equation}
and using the amplitudes (\ref{z11}) and their generalization for the QMRK,
we get for the $s$-channel imaginary part of the amplitudes 
${\cal A}_{{\cal R}}$ (details will be given elsewhere)
\[
{\cal I}{\it m}_{s}
\left(({\cal A}_{{\cal R}})_{AB}^{A^{\prime }B^{\prime }}\right) = 
\frac{s}{\left( 2\pi \right) ^{D-2}}\int \frac{d^{D-2}q_1}{\vec{q}_{1}^{~2}
\left( \vec{q}_{1}-\vec{q}\right)^{2}}\int \frac{d^{D-2}q_2}
{\vec{q}_{2}^{~2}\left( \vec{q}_{2}-\vec{q}\right) ^{2}}
\]
\begin{equation}
\times \sum_{\nu}\Phi _{A^{^{\prime
}}A}^{\left( {\cal R},\nu \right) }
\left( \vec{q}_{1};\vec{q};s_{0}\right)\int_{\delta -\infty}^{\delta+\infty} \frac{d\omega }{2\pi i}\left[ \left( \frac{s}{s_{0}}\right)
^{\omega }G_{\omega }^{\left( {\cal R}\right) }\left( \vec{q}_{1},\vec{q}%
_{2},\vec{q}\right) \right] \Phi _{B^{^{\prime }}B}^{\left( {\cal R},\nu \right) }\left( -\vec{q}_{2};
-\vec{q};s_{0}\right) 
\mbox{\ .}
\label{z15}
\end{equation}
Here $s_{0}$ is the energy scale (which can be, in principle, arbitrary), 
the index $\nu$ enumerates the states in the irreducible
representation ${\cal R}$, 
$\Phi _{P^{^{\prime }}P}^{\left( {\cal R}\mbox{,}\nu \right)}
\left( \vec{q}_{1};\vec{q};s_{0}\right)$ are the
impact factors and 
$G_{\omega }^{\left( {\cal R}\right)}
\left( \vec{q}_{1},\vec{q}_{2},\vec{q}\right)$ 
is the Mellin transform of the Green function
for the Reggeon-Reggeon scattering. The impact factors and the Green
function appear as the generalization of those defined in Refs.~[16,19] for
the case of non-forward scattering and non-vacuum quantum numbers in the $t$%
-channel. The Green function obeys the equation 
\[
\omega G_{\omega }^{\left( {\cal R}\right) }\left( \vec{q}_{1},\vec{q}_{2},%
\vec{q}\right) = 
\]
\begin{equation}
\vec{q}_{1}^{~2}\left(\vec{q}_{1}-\vec{q}\right)^{2}\delta^{\left(D-2\right)
}\left( \vec{q}_{1}-\vec{q}_{2}\right) +\int \frac{d^{D-2}{q}^{~\prime }}
{\vec{q}^{~\prime 2}\left( \vec{q}^{~\prime }-\vec{q}\right) ^{2}}{\cal K}%
^{\left( {\cal R}\right) }\left( \vec{q}_{1},\vec{q}^{~\prime };\vec{q}%
\right) G_{\omega }^{\left( {\cal R}\right) }\left( \vec{q}^{~\prime },\vec{q}%
_{2};\vec{q}\right) \mbox{\ .}  
\label{z16}
\end{equation}
Here the kernel 
\[
{\cal K}^{\left( {\cal R}\right) }\left( \vec{q}_{1},\vec{q}_{2};\vec{q}%
\right)= 
\]
\begin{equation}
\left[ \omega \left( -\vec{q}_{1}^{~2}\right) +\omega \left( -\left( 
\vec{q}_{1}-\vec{q}\right) ^{2}\right) \right]\vec{q}_{1}^{~2} \left(\vec{q}_{1}-\vec{q}\right) ^{2}
\delta ^{\left( D-2\right)
}\left( \vec{q}_{1}-\vec{q}_{2}\right) +{\cal K}_{r}^{\left( {\cal R}\right)
}\left( \vec{q}_{1},\vec{q}_{2};\vec{q}\right)  
\label{z17}
\end{equation}
is given as the sum of the ``virtual'' part, defined by the gluon
trajectory, and the ``real'' part ${\cal K}_{r}^{\left( {\cal R}\right) }$,
related to the real particle production in Reggeon-Reggeon collisions. The
``real'' part can be written in the NLLA as 
\[
{\cal K}_{r}^{\left( {\cal R}\right) }\left( \vec{q}_{1},\vec{q}_{2};\vec{q}%
\right) =\int \frac{ds_{_{RR}}}{\left( 2\pi \right)^{D}}{\cal I}{\it m}%
{\cal A}_{RR}^{\left( {\cal R}\right) }\left( q_{1},q_{2};\vec{q}\right) 
\theta \left( s_{_{\Lambda }}-s_{_{RR}}\right) 
\]
\begin{equation}
-\frac{1}{2}\int \frac{d^{D-2}q^{\prime}}{\vec{q}^{~\prime 2}\left( \vec{q}%
^{~\prime }-\vec{q}\right) ^{2}}{\cal K}_{r}^{\left( {\cal R}\right) 
\mbox{\scriptsize{B}}}
\left( \vec{q}_{1},\vec{q}^{~\prime };\vec{q}\right) 
{\cal K}_{r}^{\left( {\cal R}\right) \mbox{\scriptsize{B}}}
\left( \vec{q}^{~\prime },\vec{q}_{2};\vec{q}\right) 
\mbox{ln}\left( \frac{s_{_{\Lambda }}^{2}}{\left( \vec{q}^{~\prime }-\vec{q}%
_{1}\right) ^{2}\left( \vec{q}^{~\prime }-\vec{q}_{2}\right) ^{2}}\right) 
\mbox{\ .}  
\label{z18}
\end{equation}
In this equation 
${\cal A}^{\left( {\cal R}\right) }\left( q_{1},q_{2};\vec{q}\right) $ 
is the scattering amplitude of the Reggeons with initial momenta 
$q_{1}$ and $-q_{2}$
and momentum transfer $q$, for the representation ${\cal R}$ of the colour
group in the $t$-channel, $s_{_{RR}}=\left( q_{1}-q_{2}\right) ^{2}$ is the
squared invariant mass of the Reggeons; ${\cal K}_{r}^{\left( {\cal R}%
\right) \mbox{\scriptsize{B}}}\left( \vec{q}_{1},\vec{q}_{2};\vec{q}\right)$ 
is the part of 
the kernel at the Born (i.e. LLA) order related to the real particle 
production, which is given by the first term in the R.H.S. of Eq.~(\ref{z18}%
) taken in the Born approximation. The expression for the 
$s_{_{RR}}$-channel imaginary part 
${\cal I}{\it m}{\cal A}^{\left( {\cal R}\right) }
\left(q_{1},q_{2};\vec{q}\right)$ 
in terms of the effective vertices for the
production of particles in Reggeon-Reggeon collisions is given below. The
intermediate parameter $s_{_{\Lambda }}$ in Eq.~(\ref{z18}) must be taken
tending to infinity, so that the dependence on $s_{_{\Lambda }}$
disappears in Eq.~(\ref{z18}), because of the factorization property (\ref
{z12}) of the two gluon production vertex.

The impact factors can be expressed through the imaginary part of the
particle-Reggeon scattering amplitudes. In the NLLA the representation 
takes the form 
\[
\Phi _{P^{^{\prime }}P}^{\left( {\cal R}\mbox{,}\nu \right) }\left( \vec{q}%
_{R}\mbox{;}\vec{q}\mbox{;}s_{0}\right) =\int \frac{ds_{_{PR}}}{2\pi s}
{\cal I}{\it m}{\cal A}_{P^{^{\prime }}P}^{\left( {\cal R},\nu \right) }
\left(p_{P},q_{R};\vec{q};s_{0}\right) \theta 
\left( s_{_{\Lambda}}-s_{_{PR}}\right) 
\]
\begin{equation}
-\frac{1}{2}\int \frac{d^{D-2}q^{\prime}}{\vec{q}^{~\prime 2}\left( \vec{q}%
^{~\prime }-\vec{q}\right) ^{2}}\Phi _{P^{^{\prime }}P}^{\left( {\cal R}\mbox{%
,}\nu \right) {\mbox{\scriptsize{B}}}}\left( \vec{q}^{~\prime }\mbox{,}\vec{q}\right) 
{\cal K}_{r}^{\left( {\cal R}\right) \mbox{\scriptsize{B}}}
\left( \vec{q}^{~\prime },\vec{q}%
_{R}\right) \mbox{ln}\left( \frac{s_{_{\Lambda }}^{2}}{\left( \vec{q}%
^{~\prime }-\vec{q}_{R}\right) s_{0}}\right) \mbox{\ .}  
\label{z19}
\end{equation}
In Eq.~(\ref{z19}) $s_{_{PR}}=\left( p_{P}-q_{R}\right) ^{2}$ is the squared
particle-Reggeon invariant mass while 
${\cal I}{\it m}{\cal A}_{P^{^{\prime }}P}^{\left({\cal R},\nu \right) }
\left( p_{P},q_{R};\vec{q};s_{0}\right)$ is the 
$s_{PR}$-channel imaginary part of the scattering amplitude of the particle 
$P $ with momentum $p_{P}$ off the Reggeon with momentum $-q_{R}$, 
$\vec{q}$ being the momentum transfer. 
The argument $s_{0}$ in the impact factor
and in the amplitude shows that these two quantities depend on the
energy scale $s_{0}$ of the Mellin transformation. Of course, physical
quantities do not depend on this artificial parameter. It can be shown that
with the NLLA accuracy the R.H.S. of Eq.~(\ref{z15}) with the impact factors defined by Eq.~(\ref{z19}) and Eq.~(\ref{z27}) below, does not depend on $s_{0}$.
The Born (LLA)\ impact factors $\Phi _{P^{^{\prime }}P}^{\left( {\cal R}%
\mbox{,}\nu \right) \mbox{\scriptsize{B}}}$are given by the first term in 
the R.H.S. of Eq.~(\ref{z19}) taken in the Born approximation. 
Notice that for the Born case the integral over $s_{_{PR}}$ in 
Eq.~(\ref{z19}), as well as over $s_{_{RR\mbox{ }}}$ in Eq.~(\ref{z18}), is 
convergent, so
that the parameter $s_{_{\Lambda }}$ does not play any role. In the NLLA
the independence of the impact factors from $s_{_{\Lambda}}$ is supplied
by the factorization properties (\ref{z12}) and (\ref{z13}).

The imaginary parts of the Reggeon-Reggeon and particle-Reggeon scattering
amplitudes, entering Eqs.~(\ref{z18}) and (\ref{z19}) respectively, can be
expressed in terms of the corresponding vertices, with the help of the
operators 
$\hat{\cal {P}}_{\cal R}$
for the projection of two-gluon colour
states in the $t$-channel on the irreducible representations ${\cal R}$
of the colour group. We have 
\begin{equation}
{\cal I}{\it m}{\cal A}_{RR}^{\left( {\cal R}\right) }\left( q_{1},q_{2};
\vec{q}\right) = 
\frac{<c_{1}c_{1}^{\prime }|\hat{\cal {P}}_{\cal R}|c_{2}c_{2}^{\prime }>}
{2n_{{\cal R}}}\sum_{\left\{ f\right\} }
\int \gamma_{c_{1}c_{2}}^{\left\{ f\right\} }\left( q_{1},q_{2}\right) 
\left( \gamma_{c_{1}^{\prime }c_{2}^{\prime }}^{\left\{ f\right\} }
\left( q_{1}^{\prime},q_{2}^{\prime }\right) \right) ^{\ast }
d\rho _{f}\mbox{\ ,}  
\label{z20}
\end{equation} 
where $n_{{\cal R}}$ is the number of the states in the representation $%
{\cal R}$, $\gamma _{c_{1}c_{2}}^{\left\{ f\right\} }\left(q_{1},q_{2}\right)$ 
is the effective vertex for the production of the particles 
$\left\{ f\right\} $ in Reggeon-Reggeon collisions, $d\rho _{f}$ is their
phase space element, 
\begin{equation}
d\rho _{f}=\left( 2\pi \right)^{D}\delta^{\left( D\right) }(q_{1}-q_{2}-%
\mathop{\textstyle\sum}%
_{\left\{ f\right\} }l_{f})\prod_{\left\{ f\right\} }\frac{d^{D-1}l_{f}}{%
\left( 2\pi \right) ^{D-1}2\epsilon _{f}}\mbox{\ ,}  
\label{z21}
\end{equation}
and $q_{i}^{\prime }=q_{i}-q$. The sum over $\left\{ f\right\} 
$ in Eq.~(\ref{z20}) is performed over all the contributing particles $%
\left\{ f\right\} $ and over all their discreet quantum numbers. In the LLA
only one-gluon production does contribute; in the NLLA the contributing
states include also the two-gluon and the quark-antiquark states. The
normalization of the corresponding vertices is defined by Eq.~(\ref{z11}).

For us the most interesting representations ${\cal R}$ are the colour
singlet (vacuum) and antisymmetric colour octet (gluon) ones. We have for
the singlet case 
\begin{equation}
<c_{1}c_{1}^{\prime }|\hat{\cal {P}}_{0}|c_{2}c_{2}^{\prime }>=
\frac{\delta_{c_{1}c_{1}^{\prime }}\delta _{c_{2}c_{2}^{\prime }}}{N^{2}-1}
\mbox{\ , \ \ \ \ \ \ }n_{0}=1\mbox{\ ,}  
\label{z22}
\end{equation}
and for the octet case 
\begin{equation}
<c_{1}c_{1}^{\prime }|\hat{\cal {P}}_{8}|c_{2}c_{2}^{\prime }>=\frac{%
f_{c_{1}c_{1}^{\prime }c}f_{c_{2}c_{2}^{\prime }c}}{N}\mbox{\ , \ \ \ \ \
\ }n_{8}=8\mbox{\ ,}  
\label{z23}
\end{equation}
where $f_{abc}$ are the structure constants of the colour group. The above
matrix elements can be decomposed as
\begin{equation}
\sum_{\nu }<c_{1}c_{1}^{\prime }|\hat{\cal {P}}_{\cal R}|\nu ><\nu |%
\hat{\cal {P}}_{\cal R}|c_{2}c_{2}^{\prime }>  
\label{z24}
\end{equation}
with 
\begin{equation}
<cc^{\prime }|\hat{\cal {P}}_{0}|0>=\frac{\delta _{cc^{\prime }}}{\sqrt{N%
\mbox{ }^{2}-1}}\mbox{\ ,}  
\label{z25}
\end{equation}
\begin{equation}
<cc^{\prime }|\hat{\cal {P}}_{8}|a>=\frac{f_{acc^{\prime }}}{\sqrt{N}}
\mbox{\ .}  
\label{z26}
\end{equation}
This decomposition allows to write the imaginary part of the scattering
amplitude of the particle $P$ off the Reggeon in the form 
\[
{\cal I}{\it m}{\cal A}_{P^{^{\prime }}P}^{\left( {\cal R},\nu \right) }
\left(p_{P},q_{R};\vec{q};s_{0}\right) = 
<cc^{\prime }|\hat{\cal {P}}_{\cal R}|\nu >
\]
\begin{equation}
\times s\sum_{\left\{ f\right\}
}\int \Gamma _{\left\{ f\right\} P}^{c}(q_{R})\left( \Gamma _{\left\{
f\right\} P^{\prime }}^{c^{\prime }}(q_{R}^{\prime })\right) ^{\ast }\left( 
\frac{s_{0}}{\vec{q}_{R}^{~2}}\right) ^{\frac{1}{2}\omega \left( -\vec{q}%
_{R}^{~2}\right) }
\left( \frac{s_{0}}{\vec{q}_{R}^{~\prime 2}}\right)^{\frac{1}{2}\omega 
\left( -\vec{q}_{R}^{~\prime 2}\right) }d\rho _{f}\mbox{\ ,}
\label{z27}
\end{equation}
where $\Gamma _{\left\{ f\right\} P}^{c}$ are the effective vertices for
the production of the states $\left\{ f\right\}$. 
Their normalization is fixed by Eq.~(\ref{z11}). At the parton level
the contributing vertices are the PPR vertices $\Gamma _{P^{\prime }P}^{c}$
which have to be taken in the one-loop approximation [10,11], the vertices 
$\Gamma _{G\tilde{P}P}^{c}$ for the gluon emission in the fragmentation
region and the vertices $\Gamma _{Q\bar{Q}G}^{c}$ for the gluon $\rightarrow 
$ quark-antiquark transition \cite{FFKQ}. It is worthwhile to stress that 
for the case of the singlet representation ${\cal R}$ the
expression (\ref{z19}) for the impact factors, together with Eq.~(\ref
{z27}), is valid for colorless objects (at the hadron level) as well. For
small size objects (such as a photon with large virtuality) the impact
factors can be calculated in the perturbation theory.

\bigskip

\section{The bootstrap condition}

Let us compare the $s$-channel imaginary part of the amplitude (\ref{z3})
with the imaginary part given by Eq.~(\ref{z15}) in the case of the gluon
representation in the $t$-channel. In the LLA from Eq.~(\ref{z3}) we get 
\begin{equation}
{\cal I}{\it m}_{s}({\cal A}_{8})_{AB}^{A^{\prime }B^{\prime }}=\Gamma
_{A^{\prime }A}^{c\left( {\mbox{\scriptsize{B}}}\right) }\left( \frac{s}{|t|}\right)
^{1+\omega ^{\left( 1\right) }\left( t\right) }\pi \omega ^{\left( 1\right)
}\left( t\right) \Gamma _{B^{\prime }B\mbox{ \ \ }}^{c\left( 
{\mbox{\scriptsize{B}}}\right)}~,  
\label{z28}
\end{equation}
where the index $\left( B\right)$ denotes the Born (LLA) expression and 
$\omega ^{\left( 1\right) }$ 
stands for the gluon trajectory calculated with the 
one-loop accuracy:
\begin{equation}
\label{zz}
\omega^{(1)}(t) = \frac{g^2t}{(2\pi)^{(D-1)}}
\frac{N}{2} \int\frac{d^{D-2}k}
{\vec k^2(\vec q- \vec k)^2}~.
\end{equation}
The R.H.S. of Eq.~(\ref{z28}) coincides in the LLA with
the one of Eq.~(\ref{z15}) due to the properties of the LLA impact factors
and the Green function: 
\begin{equation}
\Phi _{P^{^{\prime }}P}^{\left( 8\mbox{,}c\right) {\mbox{\scriptsize{B}}}}=
-ig\frac{\sqrt{%
N}}{2}\Gamma _{P^{\prime }P\mbox{ \ \ }}^{c\left( {\mbox{\scriptsize{B}}}
\right)}~,
\label{z29}
\end{equation}
independently of $\vec{q}_{1}$, $\vec{q}_{2}$, and 
\begin{equation}
\int G_{\omega }^{\left( 8\right) ({\mbox{\scriptsize{B}})}}
\left( \vec{q}_{1},\vec{q}%
_{2},\vec{q}\right) \frac{d^{D-2}q_2}{\vec{q}_{2}^{~2}\left( \vec{q}_{2}-%
\vec{q}\right) ^{2}}=\frac{1}{\omega -\omega ^{\left( 1\right) }
\left(t\right) }\mbox{\ .}  
\label{z30}
\end{equation}
Applying these properties the imaginary part shown in Eq.~(\ref{z15}) in 
the NLLA becomes 
\[
{\cal I}{\it m}_{s}\left({\cal A}_{R}\right) _{AB}^{A^{^{\prime }}B^{^{\prime
}}}=\pi \left( \frac{s}{|t|}\right) ^{1+\omega ^{\left( 1\right) }\left(
t\right) }\left\{ \Gamma _{A^{\prime }A}^{c\left( {\mbox{\scriptsize{B}}}
\right) }\left[\omega^{\left( 1\right) }\left( t\right)
\left(1+\omega^{\left( 1\right) }\left( t\right)\mbox{ln}\left( \frac{|t|}{s_{0}}\right)\right)\right.\right.
\] 
\[
\left.\left. +\frac{g^{2}Nt}{2\left(2\pi \right)^{D-1}}\int \frac{d^{D-2}q_1}
{\vec{q}_{1}^{2}\left( \vec{q}_{1}-\vec{q}\right)^{2}}\int 
\frac{d^{D-2}q_2}{\vec{q}_{2}^{2}
\left( \vec{q}_{2}-\vec{q}\right) ^{2}}{\cal K}^{\left(8\right) 
\left( 1\right) }\left( \vec{q}_{1},\vec{q}_{2};\vec{q}\right) 
\mbox{ln}\left( s\right)\right]
\Gamma _{B^{\prime }B}^{c\left( {\mbox{\scriptsize{B}}}\right)}\right.
\]
\begin{equation}
\left. +ig\frac{\sqrt{N}t}{\left( 2\pi \right) ^{D-1}}
\int \frac{d^{D-2}q^{\prime}}{\vec{q}^{~\prime 2}\left( \vec{q}^{~\prime
}-\vec{q}\right) ^{2}}\left[ \Phi _{A^{^{\prime }}A}^{\left( 8,c\right)
\left( 1\right) }\left( \vec{q}^{~\prime };\vec{q};s_{0}\right) \Gamma
_{B^{\prime }B}^{c\left( {\mbox{\scriptsize{B}}}\right) }+
\Gamma _{A^{\prime }A}^{c\left( {\mbox{\scriptsize{B}}}\right) }
\Phi _{A^{^{\prime }}A}^{\left( 8,c\right) \left(1\right) }
\left( \vec{q}^{~\prime };\vec{q};s_{0}\right) \right] \right\}\mbox{\ ,}  
\label{z31}
\end{equation}
where ${\cal K}^{\left( 8\right) \left( 1\right) }$ and $\Phi
_{P^{^{\prime }}P}^{\left( 8,c\right) \left( 1\right) }$ are the
next-to-leading contributions to the kernel and to the impact factors
respectively for the gluon quantum numbers in the $t$-channel. If now we
require that this expression coincides with the imaginary part of the Regge 
form (\ref{z3}) in the NLLA: 
\[
{\cal I}{\it m}_{s}\left( {\cal A}_{R}\right)_{AB}^{A^{^{\prime }}B^{^{\prime
}}}= \pi \left( \frac{s}{|t|}\right) ^{1+\omega ^{\left( 1\right) }\left(
t\right)}\left\{ \Gamma _{A^{\prime }A}^{c\left( {\mbox{\scriptsize{B}}}
\right) }\left[\omega ^{\left( 1\right) }\left( t\right) +
\omega ^{\left( 2\right) }\left( t\right) +\omega ^{\left( 1\right) }\left(
t\right) \omega ^{\left( 2\right) }\left( t\right) \mbox{ln}\left( s\right) %
\right] \Gamma _{B^{\prime }B}^{c\left( {\mbox{\scriptsize{B}}}\right)} \right.
\]
\begin{equation}
+\left. \omega^{\left(1\right)}\left( t\right) 
\left[ \Gamma _{A^{\prime }A}^{c\left( 1\right)}
\Gamma _{B^{\prime }B}^{c\left( {\mbox{\scriptsize{B}}}\right) }+
\Gamma _{A^{\prime}A}^{c\left( {\mbox{\scriptsize{B}}}\right) }
\Gamma _{B^{\prime }B}^{c\left( 1\right) }%
\right] \right\}\mbox{\ ,}  
\label{z32}
\end{equation}
where $\Gamma _{P^{\prime }P}^{c\left( 1\right)}$ is the one-loop correction to the PPR vertex and $\omega^{\left(2\right)}\left( t\right)$ is the two-loop contribution to the trajectory, we arrive at the following bootstrap equations: 
\begin{equation}
\frac{g^{2}Nt}{2\left( 2\pi \right) ^{D-1}}\int \frac{d^{D-2}q_1}{\vec{q}%
_{1}^{~2}\left( \vec{q}_{1}-\vec{q}\right)^{2}}\int \frac{d^{D-2}q_2}{\vec{%
q}_{2}^{~2}\left( \vec{q}_{2}-\vec{q}\right)^{2}}{\cal K}^{\left(
8\right) \left( 1\right) }\left( \vec{q}_{1},\vec{q}_{2};\vec{q}\right)
=\omega ^{\left( 1\right) }\left( t\right) \omega ^{\left( 2\right) }\left(
t\right)  
\label{z33}
\end{equation}
and 
\begin{displaymath}
ig\frac{\sqrt{N}t}{\left( 2\pi \right) ^{D-1}}\int \frac{d^{D-2}q^{\prime}}{%
\vec{q}^{~\prime 2}\left( \vec{q}^{~\prime }-\vec{q}\right) ^{2}}\Phi
_{P^{^{\prime }}P}^{\left( 8,c\right) \left( 1\right) }\left( \vec{q}%
^{~\prime };\vec{q};s_{0}\right) =
\end{displaymath}
\begin{equation}
\Gamma _{P^{\prime }P}^{c\left( 1\right) }
\omega ^{\left( 1\right) }\left( t\right)
+\Gamma _{P^{\prime}P}^{c\left( {\mbox{\scriptsize{B}}}\right)}
\frac{1}{2}\left[\omega ^{\left( 2\right) }
\left( t\right)-\left(\omega ^{\left( 1\right) }
\left( t\right)\right)^2 
\mbox{ln}\left( \frac{\vec{q}^{~2}}{s_{0}}\right) \right] \mbox{\ .}
\label{z34}
\end{equation}

\bigskip

\section{Summary}
In this paper we have considered the scattering amplitudes in QCD at high
energy $\sqrt{s}$ and fixed momentum transfer $\sqrt{-t}$ in the
next-to-leading approximation in $\ln s$. Due to analyticity and crossing 
properties 
Eqs.~(\ref{z14}) and (\ref{z15}) define the 
non-forward scattering amplitudes in the NLLA in terms of the impact 
factors and the Green function for the Reggeized gluon scattering. The 
impact factors and the kernel of the equation for the Green function are 
given in terms of the gluon Regge trajectory and the Reggeon-particle 
vertices which were used also in the derivation of the  BFKL equation 
for the forward scattering in the NLLA \cite{FaLi98} and are known.

The requirement of the self consistency of the derivation of the BFKL
equation, based on the gluon Reggeization, is expressed by Eqs.~
(\ref{z33}) and (\ref{z34}). Since the BFKL equation is very important for
the theory of Regge processes at high energy $\sqrt{s}$ in perturbative
QCD, these equations should be checked. All quantities entering these equations
are unambiguously defined. The gluon Regge trajectory 
$\omega \left( t\right)$ is known with the two-loop 
accuracy \cite{FFKQ}. The one-loop correction ${\cal K}^{\left(
8\right) \left( 1\right) }$ to the kernel is given by Eqs.~(\ref{z17}), (\ref{z18}) and 
(\ref{z20}) in terms of the trajectory and the 
effective vertices for the particle production in
Reggeon-Reggeon collisions. All vertices entering these equations were
calculated with the required accuracy: the Reggeon-Reggeon-gluon vertex in
Ref.~\cite{FFLQ}, the vertices for the two-gluon production in Refs.~[9,13]
and the vertices for the quark-antiquark production in Refs.~[14,15]. The
PPR vertices entering the second bootstrap equation, i.e. Eq.~(\ref{z34}), 
were obtained with the one-loop accuracy in Refs.~[10,11]. Finally, the one
loop correction $\Phi _{P^{^{\prime }}P}^{\left( 8,c\right) \left( 1\right)
} $ to the impact factors is expressed by Eqs.~(\ref{z19}) and (\ref{z27}) 
in terms of the PPR vertices and the vertices for the production of the 
 ``excited'' states in the fragmentation 
regions. For the cases where the initial particles are quarks and gluons 
these vertices can be found in Ref.~\cite{FFKQ}.

The explicit check of the validity of Eqs.~(\ref{z33}) and (\ref{z34}) will
be the subject of subsequent publications.

\vskip 1.5cm
\underline {\bf Acknowledgment}:
One of us (V.S.F.) thanks the Dipartimento di Fisica dell'Universit\`{a} di
Milano, the Dipartimento di Fisica dell'Universit\`{a} della Calabria and
the Istituto Nazionale di Fisica Nucleare - sezione di Milano and gruppo
collegato di Cosenza for their warm hospitality during his stay in Italy. 
Fruitful discussions with G. Marchesini and M. Ciafaloni were very helpful.

\end{document}